\documentclass[11pt]{article}
\usepackage{epsfig}
\usepackage{amssymb,amsmath}
\usepackage{epigraph}

\usepackage{graphicx}				
\usepackage{amssymb}

\newcommand{\beq}{\begin{equation}}
\newcommand{\eeq}{\end{equation}}
\newcommand{\bea}{\begin{eqnarray}}
\newcommand{\eea}{\end{eqnarray}}

\begin{document}



\begin{center}

{\LARGE
The QLBS Q-Learner Goes NuQLear: 
\vskip0.5cm
 Fitted Q Iteration, Inverse RL, and Option Portfolios
}

\vskip1.0cm
{\Large Igor Halperin} \\
\vskip0.5cm
NYU Tandon School of Engineering \\
\vskip0.5cm
{\small e-mail: $igor.halperin@nyu.edu $}
\vskip0.5cm
\today \\

\vskip1.0cm
{\Large Abstract:\\}
\end{center}
\parbox[t]{\textwidth}{
The QLBS model is a discrete-time option hedging and pricing model that is based on Dynamic Programming (DP) and Reinforcement Learning (RL).
It combines the famous Q-Learning method for RL with the Black-Scholes (-Merton) model's idea of reducing the problem of option pricing and hedging to the problem of optimal rebalancing of a dynamic replicating portfolio for the option, which is made of a stock and cash.
 Here we  expand on several NuQLear (Numerical Q-Learning) topics with the QLBS model. First, we investigate the performance of Fitted Q Iteration for a RL (data-driven)
 solution to the model, and benchmark it versus a DP (model-based) solution, as well as versus the BSM model.
 Second, we develop an Inverse Reinforcement Learning (IRL) setting for the model, where we only 
 observe prices and actions (re-hedges) taken by a trader, but not rewards.
Third, we outline how the QLBS model can be used for pricing portfolios of options, rather than a single option in isolation, thus providing its own, data-driven and model independent solution to the (in)famous volatility smile problem of the Black-Scholes model. 
     
 }
 \newcounter{helpfootnote}
\setcounter{helpfootnote}{\thefootnote} 
\renewcommand{\thefootnote}{\fnsymbol{footnote}}
\setcounter{footnote}{0}
\footnotetext{
I would like to thank Eric Berger and Vivek Kapoor for stimulating discussions.
I thank Bohui Xi, Tianrui Zhao, and Yuhan Liu for an initial implementation of a DP solution of the QLBS model.
}     

 \renewcommand{\thefootnote}{\arabic{footnote}}
\setcounter{footnote}{\thehelpfootnote} 

\newpage
 
\section{Introduction}

In Ref.~\cite{IH_2017}, we presented the 
QLBS model - a discrete-time option hedging and pricing model rooted in Dynamic Programming (DP) and Reinforcement Learning (RL). 
It combines the famous Q-Learning method for RL \cite{Watkins_1989, Watkins} with the Black-Scholes (-Merton) model's idea of reducing the problem of option pricing and hedging to the problem of optimal rebalancing of a dynamic replicating portfolio for an option, which is made of a stock and cash
\cite{BS,Merton}.

In a nutshell, the celebrated Black-Scholes-Merton (BSM) model, also known as the Black-Scholes (BS) 
model \cite{BS,Merton}, shows that even though the option price can (and will) change in the future because it depends on a future stock price which is also unknown, a {\it unique}  fair option price can be found by using the principle of one price for identical goods, alongside with the method of pricing by replication. This assumes a continuous re-hedging and a special (lognormal) choice of stock price dynamics. However, such apparent uniqueness of option prices also means that, under these assumptions,  
options are completely {\it redundant}, as they can be always perfectly replicated by a simple portfolio made of a stock and cash.
 
As argued in more details in \cite{IH_2017}, an apparent redundancy of options in the BSM model is due to the fact that the latter model is formulated 
in the {\it continuous time} limit $ \Delta t \rightarrow 0 $, where hedges are rebalanced continuously, and at zero cost. In such academic limit, an option becomes risk-free, and hence completely redundant, as it is just {\it equal}, at any time $ t $, to a dynamic portfolio of a stock and cash. In any other case, i.e. when a time step $ \Delta t > 0 $,  risk in an option position cannot be completely eliminated, but at best can be {\it minimized} by a proper choice in an offsetting position in a stock that underlies the option, 
i.e. by an optimal hedge. 
 
But in the real life, re-balancing of option hedges {\it always} happens with some finite frequency $ \Delta t > 0 $, e.g. daily, monthly, etc.  
Therefore, keeping a time-step $ \Delta t $ {\it finite} while controlling {\it risk} in an option position is {\it critical} for keeping realism in {\it any} option pricing model.
While the classical BSM model gives rise to elegant closed-form expressions for option prices and hedges in the {\it mathematical} 
limit $ \Delta t \rightarrow 0 $, it makes its  theoretical {\it "risk-neutral"} option prices and hedges quite problematic in practice, even as a "zero-order" approximation to the real world.

 Indeed, as financial markets are precisely in the business of {\it trading risk}, any meaningful "zero-order" approximation should account for risk inherently present in financial options and other derivative instruments. One could argue that using an equilibrium "risk-neutral" framework for option pricing and hedging
in a risky option trading business is akin to explaining a biological system starting with equilibrium thermodynamics. While it would be absurd to describe life as a "correction" to non-life (which is the only possible state with equilibrium thermodynamics), various volatility smile models developed in continuous-time Mathematical Finance do essentially the same thing for financial risk in option pricing\footnote{"Economics ended up with the theory of
rational expectations, which maintains that there is a single optimum view of the future, that which corresponds to it, and eventually all
the market participants will converge around that view. This postulate is absurd, but it is
needed in order to allow economic theory to model itself on Newtonian Physics." (G.~Soros). I thank Vivek Kapoor for this reference.}. 

Indeed, to adjust model-based {\it "risk-neutral"} option prices to market prices of {\it risky} options, traditional 
local and/or stochastic volatility models  (see e.g. 
\cite{Wilmott}) come to the altar of Athena to ask her to breathe life into a clay {\it volatility surface} 
that was just {\it designed}  to be flat (dead) in the original 
BSM model! 
This is because the latter model rests on {\it two} critical assumptions: 1) continuous re-hedging is possible, which produces an equilibrium "risk-neutral" option price, and 2) the world is log-normal with a {\it fixed} volatility which means a {\it flat} volatility surface as a function of option strike and maturity.  Because {\it both} these assumptions are violated in practice, the original BSM model contradicts data, which makes it some way in between of a pure mathematical model, and a technical tool to quote market option prices as BS implied volatilities, and risk-manage options using their sensitivities with respect to the stock volatility ("vega"-sensitivity), and other BS sensitivity parameters ("the Greeks"). A mismatch with the market data is "fixed" by switching to local or stochastic volatility models that "match the market" much better than the original BSM model.

But this smacks of a "scientific" Cargo cult, with PDEs and GPUs replacing straw airplanes and wooden rifles.
No matter how well stochastic volatility models fit market prices, they entirely miss the {\it first} question that needs an answer for trading, namely the question of expected {\it risk} in any given option contract.
Their straight-face answer to such basic question would be "Right now, you have no risk in this option, sir!"   

Needless to say, in physics a quantum model that tweaked the Planck constant $
\hbar $ to achieve consistency with data would be deemed nonsensical, as the Planck constant is a {\it constant} that cannot change, thus any "sensitivity with respect to $ \hbar $" would be meaningless (but see \cite{Scherrer_2009}). 

Yet, a likely questionable adjustment to the original BSM model via promoting 
a {\it  model constant} (volatility) to a {\it variable} (local or stochastic volatility), to reconcile the model with market data, has become a market standard 
since 1974. The main reason for this is a common belief that advantages of analytical tractability of the classical BSM model in the continuous-time 
limit  $ \Delta t \rightarrow 0 $ outweigh its main drawbacks such as inconsistency with data, thus calling for "fixes" in the original model, such as introduction of non-constant volatilities. 
 
However, this only brings a theoretical (and practical!) nightmare on the modeling side, when the financial {\it risk}, unceremoniously thrown away in  the classical BSM model and other continuous-time models of Mathematical Finance but present in the market data, tries to make it back to the game, via mismatches between the model and market behavior. This results in what was colorfully described as "Greek tragedies" for practitioners by Satyajit Das \cite{Das}. 

The main issue with these Mathematical Finance models is that they lump together two {\it different} problems with the original BSM model: (i) the absence of risk in the limit $ \Delta t \rightarrow 0 $, 
and (ii) differences between real-world stock price dynamics and lognormal dynamics assumed in the BSM model. On the contrary, the QLBS model tackles these two problems sequentially. 

It starts with a discrete-time version of the BSM model, and re-states the problem of optimal option hedging and pricing as 
a problem of {\it risk minimization by hedging} in a sequential Markov Decision Process (MDP).
When transition probabilities and a reward function are {\it known},  such model can be solved by means of DP. 
This produces a semi-analytical solution for the option price and hedge, which only involves matrix linear algebra for a numerical implementation \cite{IH_2017}. 

On the other hand, we might know only the general {\it structure} of a MDP model, but {\it not} its specifications such as transition probability and reward function. In this case, we should solve a Bellman optimality equation for such MDP model relying only on {\it samples} of data. This is a setting 
of {\it Reinforcement Learning}, see e.g. a book by Satton and Barto \cite{SB}.  

It turns out that in such  {\it data-driven} and {\it model-free} setting, the QLBS model can be 
solved (also semi-analytically) by 
the celebrated {\it Q-Learning} method of Watkins \cite{Watkins_1989, Watkins}.  
In recognition of the fact 
that Q-Learning produces both the optimal price and optimal hedge in such
time-discretized (and distribution-free) version of the BS model, we called the model developed in Ref.~\cite{IH_2017} the QLBS model.

 While Ref.~\cite{IH_2017} focused on Mathematical Q-Learning ("MaQLear") for the QLBS model, here 
 we  expand on several topics with a Numerical Q-Learning ("NuQLear") analysis of the model. 
 First, we investigate the performance of Fitted Q Iteration (FQI) for a RL (data-driven)
 solution to the model, and benchmark it versus a DP (model-based) solution, as well as versus the BSM model.
 Second, we extend the model to a setting of Inverse Reinforcement Learning (IRL), where we only 
 observe prices and actions (re-hedges) taken by a trader, but not rewards.
Third, we outline how the QLBS model can be used for pricing portfolios of options, rather than a single option in isolation. This requires 
mutual consistency of pricing of different options in a portfolio. We show how the QLBS model addresses this problem, i.e. solves the (in)famous volatility smile problem of the Black-Scholes model. 

The paper is organized as follows. In Sect.~\ref{sect:QLBS_model}, we give a summary of the QLBS model, and present both a DP-based and RL-based solutions for the model.
An  IRL formulation for the model is developed in Sect.~\ref{sect:IRL}.
"NuQLear" experiments are presented in Sect.~\ref{sect:Experiments}.
Sect.~\ref{sect:Option_portfolios} outlines option hedging and pricing in the QLBS model in a multi-asset (portfolio) setting. 
Finally, we conclude in Sect.~\ref{sect:Summary}.     
   
\section{The QLBS model}
\label{sect:QLBS_model}

The QLBS model starts with a discrete-time version of the BSM model, where we take the view of a seller of a European option (e.g. a put option) with maturity $ T $ and a terminal payoff of $ H_T(S_T) $ at maturity, that depends on a final stock price $ S_T $ at that time. To hedge the option, the seller use the proceeds of the sale to set up a replicating (hedge) portfolio $ \Pi_t $ made of the stock $ S_t $ and a risk-free bank deposit $ B_t $. The value of the hedge portfolio at any time $ t \leq T $ is 
\beq
\label{Pi_t}
\Pi_t =  a_t  S_t + B_t
\eeq  
where $ a_t  $ is a position in the stock at time $ t $, taken to hedge risk in the option.  As at  $ t = T $ the option position should be closed, we 
set  $ u_T  = 0$, which produces a terminal condition at $ t  = T $:
\beq
\label{B_T}
\Pi_T = B_T =  H_T(S_T)
\eeq
Instead of (non-stationary) stock price $ S_t $, we prefer to use time-homogeneous variables $ X_t $ as state variables in the model, where
$ X_t $ and $ S_t $ are related as follows:
\beq
\label{X_t}
X_t = - \left( \mu - \frac{\sigma^2}{2} \right) t + \log S_t  \; \; \Leftrightarrow \; \;  S_t = e^{ X_t + \left( \mu - \frac{\sigma^2}{2} \right) t }
\eeq

\subsection{Optimal value function}

As was shown in \cite{IH_2017}, the problem of optimal option hedging and pricing in such discrete-time setting can be formulated as a problem of 
Stochastic Optimal Control (SOC) where a value function to be {\it maximized} is given by the following expression:
\beq
\label{Value_Function_port}
V_t^{\pi} ( X_t) =   \mathbb{E}_t \left[    \left. - \Pi_{t}  
-   \lambda \, \sum_{t'=t}^{T}  e^{-r (t'-t)} Var \left[  \left. \Pi_{t'}   \right| \mathcal{F}_{t'} \right]  
\right|  \mathcal{F}_t \right]  
\eeq
where $ \lambda $ is a Markowitz-like risk aversion parameter \cite{Markowitz}, $ \mathcal{F}_t $ means an information set of all Monte Carlo (or real) paths of the stock at time $ t $, and the upper-script $ \pi $ stands for a {\it policy} 
$ \pi \left(t, X_t \right) $  that maps the time $ t $ and the current state $ X_t = x_t $ into an action $ a_t \in \mathcal{A} $:
 \beq
 \label{deterministic_policy}
 a_t = \pi(t, x_t) 
\eeq 
As shown in \cite{IH_2017}, the value function (\ref{Value_Function_port}) satisfies the following Bellman equation:
\beq
\label{MDP_BSM}
V_t^{\pi}(X_t) = \mathbb{E}_{t}^{\pi} \left[ R(X_t, a_t, X_{t+1}) + \gamma V_{t+1}^{\pi} \left( X_{t+1} \right) \right]
\eeq
where the one-step time-dependent random reward is defined as follows:
\bea
\label{one-step-reward}
  R_t(X_t, a_t, X_{t+1})  &=&    \gamma a_t  \Delta S_t \left(X_t, X_{t+1} \right) - \lambda \, 
Var \left[ \left. \Pi_t  \right| \mathcal{F}_t  \right]   \nonumber  \\
 & = & \gamma a_t \Delta S_t \left(X_t, X_{t+1} \right)   -  \lambda \gamma^2  
\mathbb{E}_t \left[  \hat{\Pi}_{t+1}^2  - 2 a_t \Delta {\hat S}_t \hat{\Pi}_{t+1} + a_t^2 \left( \Delta \hat{S}_t \right)^2  \right]  
\eea
where  $ \hat{\Pi}_{t+1} \equiv \Pi_{t+1} - \bar{\Pi}_{t+1} $, 
where $ \bar{\Pi}_{t+1} $ is the sample mean of all values of  $ \Pi_{t+1} $, and similarly for $  \Delta {\hat S}_t $.
For $ t = T $, we have $ R_T =  - \lambda \, Var \left[ \Pi_T \right]  $ where $ \Pi_T $ is determined by the terminal condition 
(\ref{B_T}).

An {\it optimal policy}  $ \pi_t^{\star} (\cdot | X_t) $ is determined as a policy that maximizes the 
value function $ V_t^{\pi} \left( X_t \right) $:
\beq
\label{Value_maximization_pi}
\pi_t^{\star}(X_t) = arg \, \max_{ \pi}  \, V_t^{\pi} ( X_t)
\eeq
The optimal value function $ V_t^{\star}(X_t) $ corresponding to the optimal policy satisfies the Bellman optimality equation
\beq
\label{Bellman_V_star}
V_t^{\star}(X_t) = \mathbb{E}_{t}^{\pi^{\star}} \left[ R_t(X_t, u_t = \pi_t^{\star}(X_t), X_{t+1}) + \gamma V_{t+1}^{\star} \left( X_{t+1} \right) \right]
\eeq
Once it is solved, the (ask) option price is minus the optimal value function:   $  C_t^{(ask)} = - V_t^{\star} ( X_t) $.

If the system dynamics are known, the Bellman optimality equation can be solved using methods of Dynamic Programming such as Value Iteration. If, on the other hand, dynamics are unknown and the optimal policy should be computed using {\it samples}, which is a setting of Reinforcement Learning, then a formalism based on an action-value function, to be presented next, provides a better framework for Value Iteration methods.     
 
 \subsection{Action-value function}

The action-value function, or Q-function, is defined by an expectation of the same expression as in the definition of the value function (\ref{Value_Function_port}), but 
 conditioned on both the current state $ X_t $ {\it and} the initial action $ a = a_t $, while following a policy $ \pi $ afterwards:  
\bea
\label{Value_maximization}
& Q_t^{\pi} (x,a) = 
  \mathbb{E}_t \left[ \left. - \Pi_{t}(X_t) \right| X_t = x, a_t = a \right]     \\
&-   \lambda \,  \mathbb{E}_t^{\pi} \left[  \left. \sum_{t'=t}^{T}  e^{-r (t'-t)} Var \left[  \left. \Pi_{t'}(X_{t'} )   \right| \mathcal{F}_{t'} \right]  
\right|  X_t = x, a_t = a \right] \nonumber 
\eea
The Bellman equation for the Q-function reads \cite{IH_2017}
\beq
\label{Bellman_Q_1}
Q_t^{\pi}(x, a) = \mathbb{E}_{t} \left[ \left. R_t(X_t, a_t, X_{t+1})  \right| X_t = x, a_t = a \right] 
+ \gamma \mathbb{E}_{t}^{\pi} \left[ \left. V_{t+1}^{\pi} \left( X_{t+1} \right) \right| X_t = x \right]
\eeq
An optimal action-value function $ Q_T^{\star} (x,a) $ is obtained when (\ref{Value_maximization}) is evaluated with an optimal policy 
$ \pi_t^{\star} $:
\beq
\label{opt_policy_Q}
\pi_t^{\star} = \arg \, \max_{\pi} Q_t^{\pi} (x,a)
\eeq
The optimal value- and state-value functions are connected by the following equations
\bea
\label{V_A}
& & V_t^{\star}(x) = \max_{a} Q_t^{\star} (x,a)  \\
& & Q_t^{\star} (x,a) =  \mathbb{E}_t \left[   R_t(x, a, X_{t+1})  \right] 
+ \gamma \mathbb{E} \left[ \left. V_{t+1}^{\star}(X_{t+1})  \right| X_t = x 
\right]  \nonumber 
\eea
The Bellman Optimality equation for the action-value function is obtained by substituting the first of Eqs.(\ref{V_A}) into the second one:
\beq
\label{Bellman_Q}
Q_t^{\star} (x,a) =  \mathbb{E}_t \left[   R_t \left(X_t, a_t, X_{t+1} \right)   + \gamma \max_{a_{t+1} \in \mathcal{A}}  
 \left. Q_{t+1}^{\star} \left( X_{t+1}, a_{t+1} \right)  \right| X_t = x, a_t = a  \right]  \, , \; \; t = 0, \ldots, T-1
\eeq
with a terminal condition at $ t = T $ given by 
 \beq
 \label{Q_T}
 Q_T^{\star}(X_T, a_T = 0) = - \Pi_T \left(X_T \right) - \lambda \, Var \left[ \Pi_T \left( X_T \right) \right]
 \eeq 
 where $ \Pi_T $ is determined by the terminal condition (\ref{B_T}). 
A "greedy" policy $ \pi^{\star} $ that is used in the QLBS model always seeks an action that maximizes the action-value function in the 
current state:
\beq
\label{pi_greedy}
\pi_t^{\star} (X_t) = arg \, \max_{a_t \in \mathcal{A}} Q_t^{\star} (X_t,a_t)
\eeq

\subsection{DP solution for the optimal Q-function}
\label{sect:DP}

If transition probabilities to compute the expectation in the right-hand side of the Bellman optimality equation (\ref{Bellman_Q}) are {\it known}, then the Bellman equation (\ref{Bellman_Q}) can be solved, jointly with the optimal policy (\ref{pi_greedy}), using backward recursion starting from $ t = T-1 $ and the terminal condition (\ref{Q_T}). 
This can be used for benchmarking in our test environment where we {\it do} know these probabilities, and know the rewards function (\ref{one-step-reward}).

Substituting the one-step reward (\ref{one-step-reward}) into the Bellman optimality equation 
(\ref{Bellman_Q})  we find that $ Q_t^{\star} \left( X_t, a_t \right) $ is {\it quadratic} in the action variable  $ a_t $:
\bea
\label{Bellman_Q_a_t}
  Q_{t}^{\star}(X_{t},a_{t}) &=& \gamma \mathbb{E}_{t} \left[ 
 Q_{t+1}^{\star} \left( X_{t+1}, a_{t+1}^{\star} \right) 
 + a_t \Delta S_t \right]   \nonumber \\
 &- &   \lambda \gamma^2   \, \mathbb{E}_t  \left[ \hat{\Pi}_{t+1}^{2}  - 2  a_t  \hat{\Pi}_{t+1} \Delta \hat{S}_t
 + a_t^2  \left( \Delta \hat{S}_t \right)^2  \right] \, , \; \; t = 0, \ldots, T-1
\eea
 As $ Q_t^{\star} \left( X_t, a_t \right) $  is a quadratic function of $ a_t $, the optimal action (i.e. the hedge) $ a_t^{\star} (S_t) $ that maximizes 
$ Q_{t}^{\star}(X_{t},a_{t}) $  is computed analytically:
\beq
\label{a_star_t}
a_{t}^{\star} \left( X_t \right)
 =  \frac{\mathbb{E}_{t} \left[  \Delta \hat{S}_{t}  \hat{\Pi}_{t+1} + \frac{1}{2 \gamma \lambda} \Delta S_{t} \right]}{
  \mathbb{E}_{t} \left[ \left( \Delta \hat{S}_{t} \right)^2 \right]} 
\eeq 
Plugging Eq.(\ref{a_star_t}) back into Eq.(\ref{Bellman_Q_a_t}), we obtain an explicit 
recursive formula for the {\it optimal} action-value function:
\beq
\label{Q_star_rec}
Q_{t}^{\star}(X_{t},a_{t}^{\star}) = \gamma \mathbb{E}_{t} \left[ Q_{t+1}^{\star}(X_{t+1},a_{t+1}^{\star}) - \lambda \gamma 
\hat{\Pi}_{t+1}^2 + \lambda \gamma  \left( a_{t}^{\star} \left(X_t \right) \right)^2   \left( \Delta \hat{S}_{t} \right)^2  \right] 
 \, , \; \; t = 0, \ldots, T-1
\eeq
where $ a_{t}^{\star} \left( X_t \right) $ is defined in Eq.(\ref{a_star_t}). 
 
In practice, the backward recursion expressed by Eqs.(\ref{Q_star_rec}) and (\ref{a_star_t}) is solved in a Monte Carlo setting, where we assume to have access to 
$ N_{MC} $ simulated (or real) paths for the state variable $ X_t $ \cite{IH_2017}. 
In addition, we  assume that we have chosen a set of basis functions  $ \{ \Phi_n(x) \} $.

We can then expand  the optimal action (hedge)  $ a_t^{\star} \left( X_t \right)  $ and 
optimal Q-function $ Q_t^{\star} \left(X_t, a_t^{\star} \right) $ 
 in basis functions, with time-dependent coefficients:
\beq
\label{Q_basis_exp}
a_t^{\star} \left( X_t \right) = \sum_{n}^{M}  \phi_{nt} \Phi_n \left( X_t  \right) \, ,  \; \;  
Q_t^{\star} \left(X_t, a_t^{\star} \right) = \sum_{n}^{M} \omega_{nt} \Phi_n \left( X_t  \right)  
\eeq
Coefficients $ \phi_{nt} $ and $ \omega_{nt} $ are computed recursively backward in time for $ t = T-1, \ldots, 0 $.
The results are given by the following expressions:
\beq
\label{phi_nt_vec}
{\bf \phi}_t^{\star} = {\bf A}_t^{-1} {\bf B}_t
\eeq 
where 
\bea
\label{AB}
A_{nm}^{(t)} &=& \sum_{k=1}^{N_{MC}} \Phi_n \left( X_t^k \right)  \Phi_m \left( X_t^k \right) \left(  \Delta \hat{S}_t^k  \right)^2 
\nonumber \\
B_{n}^{(t)} &=& \sum_{k=1}^{N_{MC}} \Phi_n \left( X_t^k \right)  \left[ \hat{\Pi}_ {t+1}^k \Delta \hat{S}_t^k
+ \frac{1}{2 \gamma \lambda} \Delta S_t^k  \right]
\eea
and 
\beq
\label{omega_nt_vec}
\omega_{t}^{\star} =  {\bf C}_t^{-1} {\bf D}_t
\eeq
where 
\bea
\label{CD}
C_{nm}^{(t)} &=& \sum_{k=1}^{N_{MC}} \Phi_n \left( X_t^k \right)  \Phi_m \left( X_t^k \right)
\nonumber \\
D_{n}^{(t)} &=& \sum_{k=1}^{N_{MC}} \Phi_n \left( X_t^k \right)  
 \left(   R_t \left(X_t^k, a_t^{k \star}, X_{t+1}^k \right)   + \gamma \max_{a_{t+1} \in \mathcal{A}}  
  Q_{t+1}^{\star} \left( X_{t+1}^k, a_{t+1} \right)   \right)
\eea

Equations (\ref{phi_nt_vec}) and (\ref{omega_nt_vec}), computed jointly and recursively for $ t = T-1, \ldots, 0 $,
provide a practical implementation of the DP-based solution to the QLBS model using expansions in basis functions. This approach can be used to find optimal price and optimal hedge when the dynamics are {\it known}. For more details, see Ref.~\cite{IH_2017}.

\subsection{RL solution for QLBS: Fitted Q Iteration}
\label{sect:FQI}

Reinforcement Learning (RL) solves the same problem as Dynamic Programming (DP), i.e. it finds an optimal policy.
But unlike DP, RL does {\it not} assume that transition probabilities and reward function are known.
Instead, it relies on {\it samples} to find an optimal policy. 

Our setting assumes a {\it batch-mode} learning, when we only have access to some historically collected data.
The data available is given by a set of $ N_{MC} $ trajectories for the underlying stock $ S_t $ (expressed as a function of $ X_t $ 
using Eq.(\ref{X_t})), hedge position $ a_t $ , instantaneous reward $ R_t $, and the next-time value $ X_{t+1} $:
\beq
\label{F_t_RL}
\mathcal{F}_t^{(n)} = \left\{  \left( X_t^{(n)}, a_t^{(n)}, R_t^{(n)}, X_{t+1}^{(n)} \right) \right\}_{t=0}^{T-1} \, ,  \; \; n = 1, \ldots, N_{MC} 
\eeq
We assume that such dataset is available either 
as a simulated data, or as a real historical stock price data, combined with real trading data or artificial data that would track the 
performance of a hypothetical stock-and-cash replicating portfolio for a given option.

We use a popular batch-model Q-Learning method called Fitted Q Iteration (FQI) \cite{Ernst, Murphy}. A starting point in this method 
is a choice of a parametric family of models for quantities of interest, namely optimal action and optimal action-value function. We use linear architectures where
functions sought are {\it linear} in adjustable parameters that are next optimized to find the optimal action and action-value function. 

We use the same set of basis functions  $ \{ \Phi_n(x) \} $ as we used above in Sect.~\ref{sect:DP} .
As the optimal Q-function $ Q_t^{\star} \left(X_t, a_t \right) $ is a quadratic function of $ a_t $, we can represent 
it as an expansion  
 in basis functions, with time-dependent coefficients parametrized by a matrix $ {\bf W}_t $:
\bea
\label{Q_any_a}
Q_t^{\star} \left(X_t, a_t \right) &=&
\left( 1, a_t, \frac{1}{2} a_t^2 \right) 
 \, \left( \begin{array}{cccc}
W_{11}(t)  & W_{12}(t) & \cdots & W_{1M}(t)   \\
W_{21}(t)  & W_{22}(t) & \cdots  & W_{2M}(t) \\  	
W_{31} (t) & W_{32}(t) & \cdots & W_{3M} (t)   \\
\end{array} \right)  
\left( \begin{array}{c}
\Phi_1(X_t)  \\
\vdots \\
\Phi_M(X_t)   \\
\end{array} \right) 
 \nonumber \\ 
&\equiv & {\bf A}_t^T   {\bf W}_t  {\bf \Phi}(X_t)
\equiv   {\bf A}_t^T  \, {\bf U}_{W} (t,X_t)  
\eea 
Eq.(\ref{Q_any_a}) is further re-arranged to convert it into a product of a parameter vector and a vector that 
depends on both the state and the action:    
\bea
\label{rearrange}
 Q_t^{\star} \left(X_t, a_t \right)  &=& {\bf A}_t^T   {\bf W}_t  {\bf \Phi}(X) = 
 \sum_{i=1}^{3} \sum_{j=1}^{M} \left( {\bf W}_t \odot  \left( {\bf A}_t  \otimes  {\bf \Phi}^T(X)  
 \right) \right)_{ij}  \nonumber \\
 &=& \vec{{\bf W}}_t \cdot vec 
 \left( {\bf A}_t  \otimes
 {\bf \Phi}^T(X) \right) \equiv  \vec{{\bf W}}_t \vec{{\bf \Psi}} \left(X_t,a_t \right)
 \eea
Here $ \odot $ stands for an element-wise (Hadamard) product of two matrices. The vector of time-dependent parameters 
$  \vec{{\bf W}_t}  $ is obtained by concatenating columns of matrix $ \bf{W}_t $, and similarly, 
$ \vec{{\bf \Psi}} \left(X_t,a_t \right) = 
 vec \left( {\bf A}_t  \otimes {\bf \Phi}^T(X) \right) $ stands for 
a vector obtained by concatenating columns of the outer product of vectors $ {\bf A}_t $ and $ {\bf \Phi}(X) $.
 
Coefficients  $  \vec{\bf{W}}_t  $ can then be computed recursively backward in time for $ t = T-1, \ldots, 0 $ \cite{IH_2017}:
\beq
\label{W_opt_vec}
 \vec{\bf{W}}_{t}^{\star} =  {\bf S}_t^{-1} {\bf M}_t
\eeq
where
\bea
\label{SM}
S_{nm}^{(t)} &=& \sum_{k=1}^{N_{MC}} \Psi_n \left( X_t^k, a_t^k \right)  \Psi_m \left( X_t^k, a_t^k \right)
\nonumber \\
M_{n}^{(t)} &=& \sum_{k=1}^{N_{MC}} \Psi_n  \left( X_t^k, a_t^k \right)  
 \left(   R_t \left(X_t^k, a_t^k, X_{t+1}^k \right)   + \gamma \max_{a_{t+1} \in \mathcal{A}}  
  Q_{t+1}^{\star} \left( X_{t+1}^k, a_{t+1} \right)   \right)
\eea
To perform the maximization step in the second equation in (\ref{SM}) analytically, note that 
because coefficients ${\bf W}_{t+1} $ and hence vectors $  {\bf U}_{W} (t+1, X_{t+1})  \equiv  {\bf W}_{t+1} {\bf \Phi}(X_{t+1})
$ (see Eq.(\ref{Q_any_a})) are known from the previous step, we have
\beq
\label{max_Q}
Q_{t+1}^{\star} \left( X_{t+1}, a_{t+1}^{\star}  \right) =   \mathbf U_W^{\left(0\right)} \left(t+1,X_{t+1} \right) +
a_{t+1}^{\star} \mathbf U_W^{\left(1\right)} \left(t+1,X_{t+1} \right) + 
 \frac{\left( a_{t+1}^{\star} \right)^2 }{2} \mathbf U_W^{\left(2\right)} \left(t+1,X_{t+1} \right) 
\eeq
It is important to stress here that while this is a quadratic expression in $ a_{t+1}^{\star} $, it would be completely {\it wrong} to use a point of its maximum 
as a function of  $ a_{t+1}^{\star} $ as such optimal value in Eq.(\ref{max_Q}).  
This would amount to using the same dataset to estimate both the optimal action and the optimal Q-function, leading to an overestimation of  $ Q_{t+1}^{\star} \left( X_{t+1}, a_{t+1}^{\star}  \right)  $ in Eq.(\ref{SM}), due to Jensen's inequality and convexity of the $ \max(\cdot ) $ function.
 The correct way to use Eq.(\ref{max_Q}) is to plug there a value of $ a_{t+1}^{\star} $ computed using the analytical solution 
 Eq.(\ref{a_star_t}), applied at the previous time step. Due to availability of the analytical optimal action  (\ref{a_star_t}),
 a potential overestimation problem, a classical problem of Q-Learning that is sometimes addressed using such methods as Double Q-Learning 
 \cite{Double-Q}, is avoided in the QLBS model, leading to numerically stable results.

Equation (\ref{W_opt_vec}) gives the solution for the QLBS model in a {\it model-free} and {\it off-policy} setting,
via its reliance on  Fitted Q Iteration which {\it is} a model-free and off-policy algorithm \cite{Ernst, Murphy}.

\section{Inverse Reinforcement Learning in QLBS}
\label{sect:IRL}

Inverse Reinforcement Learning (IRL) provides a very interesting and useful extension of the (direct) RL paradigm. In the context of batch-mode
learning used in this paper, a setting of IRL is nearly identical to the setting of RL (see Eq.(\ref{F_t_RL})), except that there is no information about rewards:
\beq
\label{F_t_IRL}
\mathcal{F}_t^{(n)} = \left\{  \left( X_t^{(n)}, a_t^{(n)}, X_{t+1}^{(n)} \right) \right\}_{t=0}^{T-1} \, ,  \; \; n = 1, \ldots, N 
\eeq
The objective of IRL is typically two-fold: (i) find rewards $  R_t^{(n)} $ that would be most consistent with observed states and action, and (ii) (the same as in RL) find the optimal policy and action-value function. One can distinguish between {\it on-policy} IRL and {\it off-policy} IRL. In the former case, we
know that observed actions were {\it optimal} actions. In the latter case, observed actions may {\it not} necessarily follow an optimal policy, and can be 
sub-optimal or noisy. 

In general, IRL is a harder problem than RL. Indeed, not only we have to find optimal policy from data, which is the same task as in RL, but we also have to do it without observing rewards. Furthermore, the other task of IRL is to find a (the?) reward function corresponding to an observed sequence of states and actions. Note that situations with missing reward information are probably encountered more frequently in potential applications  of RL/IRL than  
observable rewards. In particular, this is typically the case when RL methods are applied  to study human behavior, see e.g. \cite{Krishnan}. IRL is also widely used in robotics as a useful alternative to direct RL methods via training robots by demonstrations, see e.g. \cite{Kober}. 

It appears that IRL offers a very attractive, at least conceptually, approach for many financial applications that consider rational agents involved in a sequential decision process, where no information about rewards received by an agent is available to a researcher. Some examples of such (semi- ?) rational agents would be loan or mortgage borrowers, deposit or saving account holders, credit card holders, consumers of utilities such as cloud computing, mobile data, electricity, etc.

In the context of trading applications, such IRL setting may arise when a trader wants to learn a strategy of a counterparty. 
She observes counterparty's actions in their bilateral trades, but not counterparty's rewards.
Clearly, if she reverse-engineered most likely counterparty's rewards from observed actions to find counterparty's objective (strategy), she could use it to design her own strategy. This is a typical IRL problem.

While typically IRL is a harder problem than RL, and {\it both} are computationally {\it hard}, in the QLBS model both are about equally {\it easy}, due to 
a quadratic form of both the reward function (\ref{one-step-reward}) and action-value function (\ref{Bellman_Q_a_t}). Moreover, the general IRL setting, where only states and actions, but not rewards, are observed in a dataset, is exactly in between of our two previous settings: a DP setting where we only observe states, and a RL setting where we observe states, actions, and rewards. 

The main difference is that in the DP setting we know model dynamics, including in particular the risk aversion parameter $ \lambda $, while in
the setting of RL or IRL  $ \lambda $ is unknown. Therefore, we will first assume that $ \lambda $ is {\it known}, and outline how IRL should proceed with the QLBS model, and then we will discuss ways to estimate $ \lambda $ from data.  

In the IRL setting, once we have observed states $ X_t $ and actions $ a_t $, rewards $ R_t $ corresponding to these actions can be obtained, 
if $ \lambda $ is known, in  
the same way they were computed in  Eq.(\ref{one-step-reward}). The only difference is that while 
in the DP solution of Sec.~\ref{sect:DP} we computed rewards  (\ref{one-step-reward}) for {\it optimal} actions (\ref{a_star_t}), in the IRL setting we would use
{\it observed} actions $ a_t $ to plug into  Eq.(\ref{one-step-reward}) to compute the corresponding rewards. After that, 
the algorithm proceeds in the same way as the FQI solution of Sect.~\ref{sect:FQI}, using these computed rewards instead of observed rewards
in Eq.(\ref{SM}). Clearly, this produces {\it identical} RL and IRL solutions of the QLBS model, as long as $ \lambda $ implied in observed rewards 
$ R_t $ in the RL case is the same $ \lambda $ used in Eq.(\ref{one-step-reward}) by the IRL solution.

This means that the first problem of IRL, i.e. finding a reward function, amounts for the QLBS model to finding just {\it one} parameter $ \lambda $ 
using Eq.(\ref{one-step-reward}). This can be done using an approach that we present next.

\subsection{Maximum Entropy IRL}

A simple method  to estimate the one-step reward function  (\ref{one-step-reward}) by estimating its parameter $ \lambda $ is based on a highly tractable version of a popular Maximum Entropy (MaxEnt) IRL method \cite{Ziebart_2008} that was developed in \cite{IRL_marketing} in a different context. 

We start with writing expected rewards corresponding to Eq.(\ref{one-step-reward}) as follows
\beq
\label{R_t}
 \bar{R}_t(X_t, a_t)  \equiv \mathbb{E}_t \left[  R_t(X_t, a_t, X_{t+1})  \right]  =   c_0 (\lambda) + a_t  c_1 (\lambda) - \frac{1}{2} a_t^2  c_2 (\lambda)
\eeq
where, omitting for brevity the dependence on $ X_t $, we defined
\beq
\label{reward_coeffs}
 c_0 (\lambda) =   -  \lambda \gamma^2   \mathbb{E}_t \left[  \hat{\Pi}_{t+1}^2  \right]  \, , \; 
 c_1 (\lambda) =  \gamma   \mathbb{E}_t \left[ \Delta S_t  + 2 \lambda \gamma  \Delta {\hat S}_t \hat{\Pi}_{t+1}  \right]  
 \, , \; 
 c_2(\lambda) = 2 \lambda \gamma^2  \mathbb{E}_t \left[ \left( \Delta \hat{S}_t \right)^2  \right]  
\eeq 
The MaxEnt method of \cite{IRL_marketing} assumes that one-step probabilities of observing different actions $ a_t $ in data are described by an exponential model
\beq
\label{MaxEnt}
p_{\lambda} \left( \left. a_t \right| X_t \right) = \frac{1}{Z_{\lambda}} e^{  \bar{R}_t(X_t, a_t) } = \sqrt{ \frac{c_2(\lambda)}{ 2 \pi} }   
\exp \left[ - \frac{c_2(\lambda)}{2}  \left( a_t - \frac{c_1 (\lambda)}{c_2 (\lambda)} \right)^2 \right]
\eeq 
where $ Z_\lambda $ is a normalization factor. 

Thus, by combining an exponential distribution of the MaxEnt method with the quadratic expected reward (\ref{R_t}), we ended up with a Gaussian 
action distribution (\ref{MaxEnt}) for IRL in QLBS. Clearly this is very good news given the amount of tractability of Gaussian distributions.  
Using Eq.(\ref{MaxEnt}), the log-likelihood of observing data $ 
\left\{ X_t^{(k)}, a_t^{(k)} \right\})_{k=1}^{N} $ is (omitting a constant factor $ - \frac{1}{2} \log \left( 2 \pi \right)  $ in the second expression)    
\beq
\label{ll}
LL (\lambda) = \log \prod_{k=1}^{N} p_{\lambda} \left( \left. a_t^{(k)} \right| X_t^{(k)} \right) =  \sum_{k=1}^{N} \left( \frac{1}{2} \log c_2^{(k)}(\lambda)  
- \frac{c_2^{(k)}(\lambda)}{2} \left( a_t^{(k)} - \frac{c_1^{(k)}(\lambda)}{  c_2^{(k)}(\lambda)} \right)^2 \right)
\eeq
where $ c_i^{(k)}(\lambda) $ with $ i = 1,2 $ stands for expressions (\ref{reward_coeffs}) evaluated on the $k$-th path. As this is a concave function 
of $ \lambda $, its unique maximum can be easily found numerically using standard optimization packages. 

Note that optimization in Eq.(\ref{ll}) refers to one particular value of $ t $, therefore this calculation can be repeated independently for different 
times $ t $, producing a curve $ \lambda_{impl}(t) $ that could be viewed as a term structure of implied risk aversion parameter.

It can also be noticed that while Eq.(\ref{MaxEnt}) describes a {\it probabilistic} Gaussian policy (action probability), in Sect.~\ref{sect:FQI} we used the 
{\it deterministic} 
"greedy" policy (\ref{pi_greedy}). Therefore,  if we used a value of $ \lambda $ estimated with Eq.(\ref{ll}) in the IRL algorithm described above,  this may not produce the same result as the RL approach of Sect.~\ref{sect:FQI}. Policy assumptions can be made more consistent between the RL and IRL 
approaches if instead of Q-Learning (in the form of Fitted Q Iteration) that we used in Sect.~\ref{sect:FQI}, we switched to G-Learning \cite{G-Learning} that replaces
the "greedy max" term in Eq.(\ref{SM}) with a "soft-greedy max" term for a G-function:
\beq
\label{G-learning}
\max_{a_{t+1} \in \mathcal{A}}  
  Q_{t+1}^{\star} \left( X_{t+1}, a_{t+1} \right)  \rightarrow - \frac{1}{\beta} \log \left( \int p \left(a | X_{t+1} \right) 
  e^{- \beta G_{t+1} \left(X_{t+1}, a \right)} da \right)
 \eeq
 where $ \beta $ is an "inverse temperature" parameter of G-Learning \cite{G-Learning}. We leave G-Learning in the QLBS model for a future research.
   
\section{NuQLear experiments}
\label{sect:Experiments}

We illustrate the numerical performance of the model in different settings (DP, RL, IRL) using simulated stock price histories $ S_t $ with the initial stock price $ S_0 = 100 $, stock drift $ \mu = 0.05 $, and volatility $ \sigma = 0.15 $. Option maturity is $ T = 1 $ year, and a risk-free rate is $ r = 0.03 $.
We consider an ATM ("at-the-money") European put option with strike $ K = 100 $. Re-hedges are done bi-weekly (i.e. $ \Delta t = 1/24 $).  
We use $ N_{MC} = 50,000 $ Monte Carlo scenarios for a path of the stock,
and report results obtained with two MC runs, where the error reported is equal to one standard deviation calculated from these runs.
In our experiments, we use pure risk-based hedges, i.e. 
omit the second term in Eq.(\ref{a_star_t}) to facilitate comparison with the BSM model.

We use 12 basis functions chosen to be cubic B-splines on a range of values of $ X_t $ between the smallest and largest values observed in a dataset.  

\subsection{DP solution}

\begin{figure}[ht]
\begin{center}
\includegraphics[
width=90mm,
height=70mm]{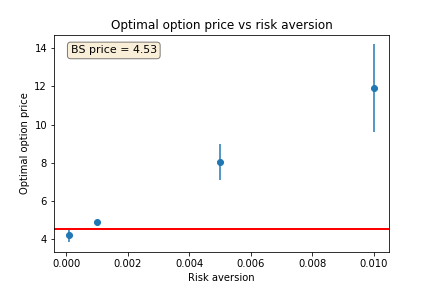}
\caption{The ATM put option price vs risk aversion parameter. The horizontal red line corresponds to the BS model price. Error bars correspond to one standard deviation of two MC runs.} 
\label{fig:price_vs_lambda}
\end{center}
\end{figure}

In our experiments below, we pick the Markowitz risk aversion parameter $ \lambda = 0.001 $. This provides a visible difference of QLBS prices from BS prices, while being not too far away from BS prices. The dependence of the ATM option price on $ \lambda $ is shown in Fig.~\ref{fig:price_vs_lambda}.
   
\begin{figure}[ht]
\begin{center}
\includegraphics[
width=140mm,
height=120mm]{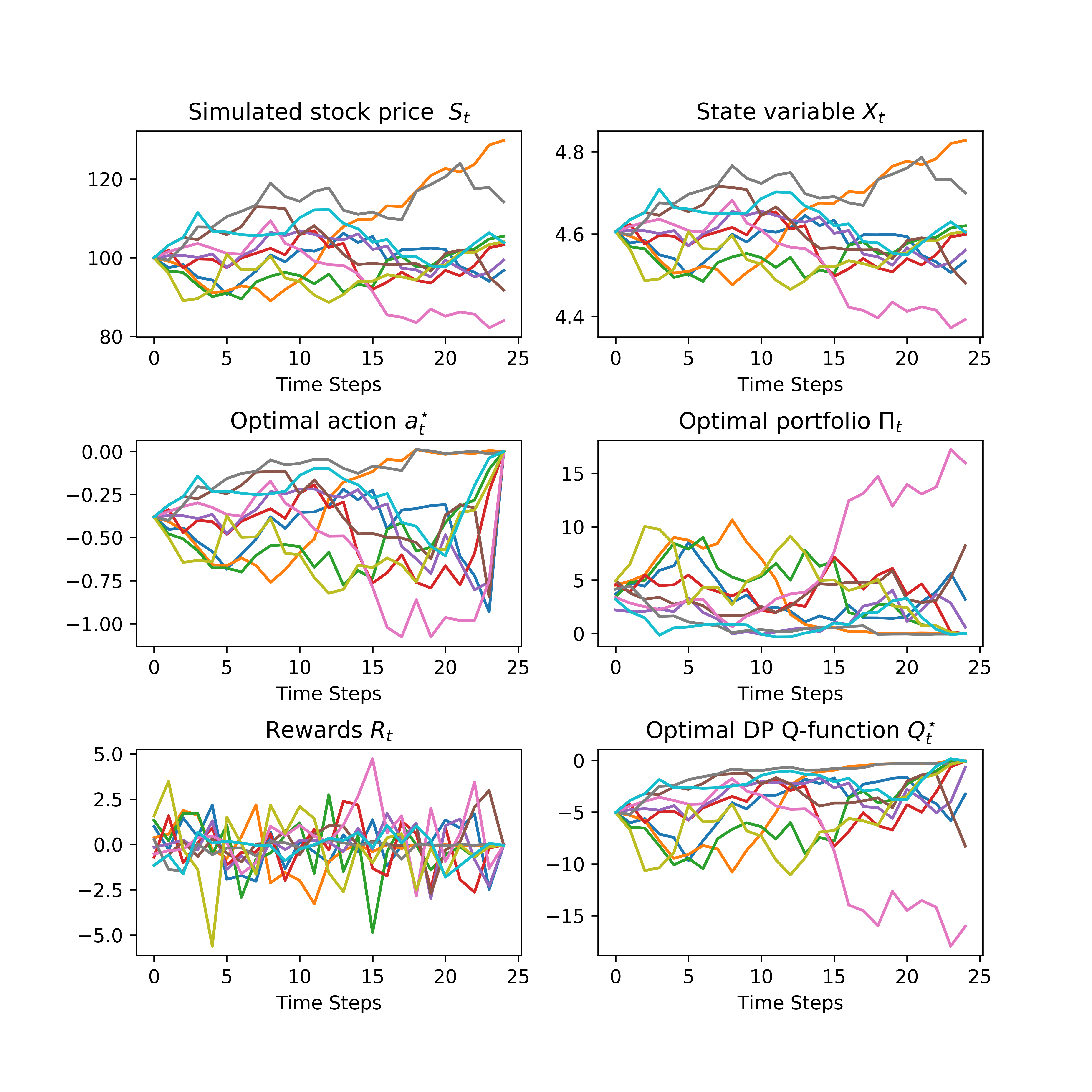}
\caption{DP solution for the ATM put option on a sub-set of MC paths} 
\label{fig:QLBS_DP_summary_graphs_ATM_option_mu_r}
\end{center}
\end{figure}

Simulated path and solutions for optimal hedges, portfolio values, and Q-function values corresponding to the DP solution of Sect.~\ref{sect:DP} are illustrated in Fig.~\ref{fig:QLBS_DP_summary_graphs_ATM_option_mu_r}. In the numerical implementation of matrix inversion in 
Eqs.(\ref{phi_nt_vec}) and (\ref{omega_nt_vec}), we used a regularization by adding a unit matrix with a regularization parameter of $ 10^{-3} $.  

The resulting QLBS ATM put option price is $  4.90 \pm 0.12  $ (based on two MC runs), while the BS price is 4.53. 

\subsection{On-policy RL/IRL solutions}

\begin{figure}[ht]
\begin{center}
\includegraphics[
width=140mm,
height=120mm]{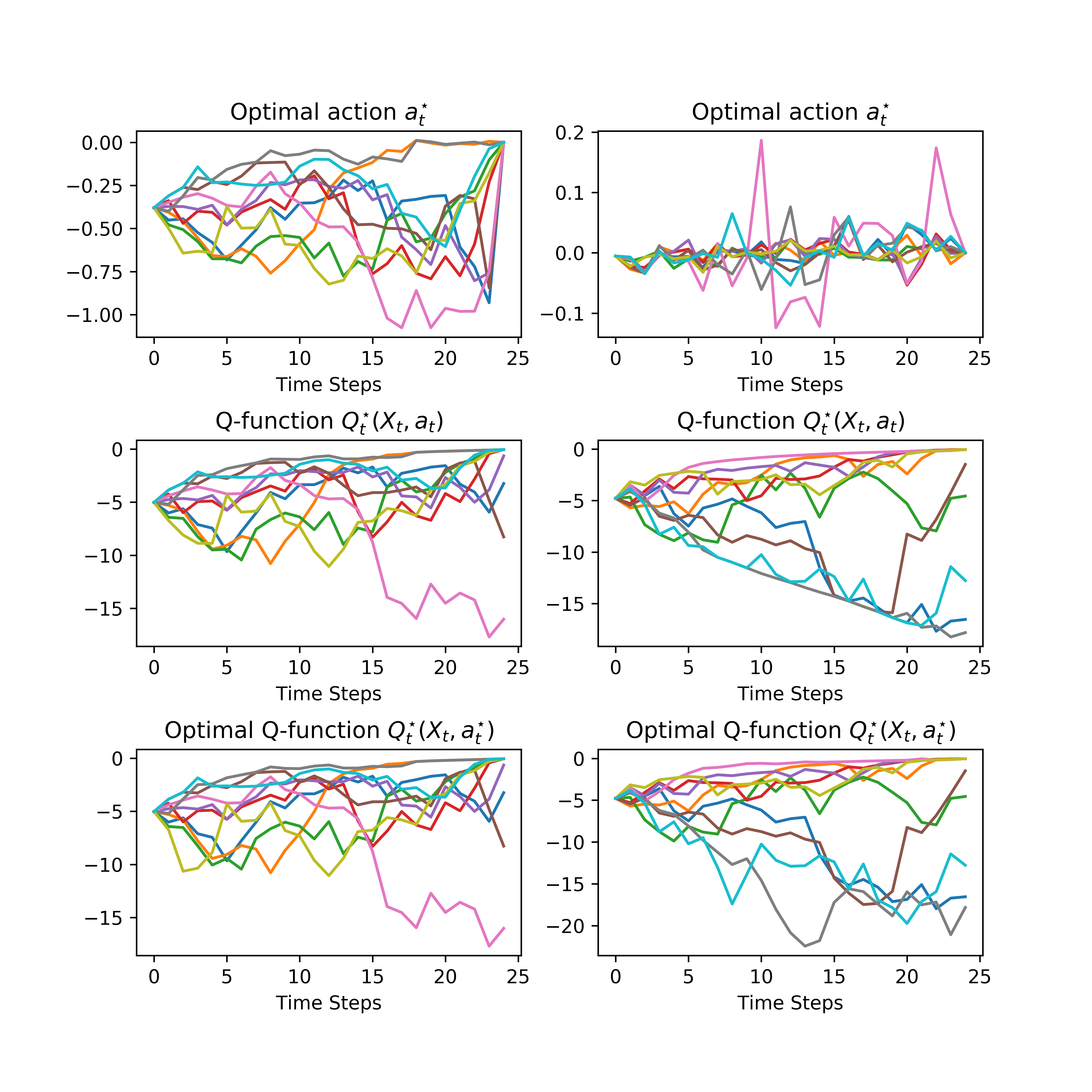}
\caption{RL solution (Fitted Q Iteration) for {\it on-policy} learning for the ATM put option on a sub-set of MC paths for two MC runs.} 
\label{fig:QLBS_FQI_summary_graphs_ATM_option_mu_r}
\end{center}
\end{figure}

We first report results obtained with {\it on-policy} learning. In this case, optimal actions and rewards computed as a part of a DP solution are used as inputs to the Fitted Q Iteration algorithm Sect.~\ref{sect:FQI} and the IRL method of Sect.~\ref{sect:IRL}, in addition to the paths of the underlying stock. 
Results of two MC runs with Fitted Q Iteration algorithm of Sect.~\ref{sect:FQI} are shown in  
Fig.~\ref{fig:QLBS_FQI_summary_graphs_ATM_option_mu_r}. Similarly to the DP solution, we add a unit matrix with a regularization parameter of $ 10^{-3} $ to invert matrix $ {\bf C}_t $ in Eq.(\ref{W_opt_vec}). Note that because here we deal with {\it on-policy} learning, the resulting optimal Q-function
$  Q_t^{\star} \left(X_t, a_t \right) $ and its optimal value $  Q_t^{\star} \left(X_t, a_t^{\star} \right) $ are virtually identical in the graph.
The resulting QLBS RL put price is $ 4.90 \pm 0.12 $ which is identical to the DP value. As expected, the IRL method of Sect.~\ref{sect:IRL} produces 
the same result.

\subsection{Off-policy RL solution}

\begin{figure}[ht]
\begin{center}
\includegraphics[
width=120mm,
height=90mm]{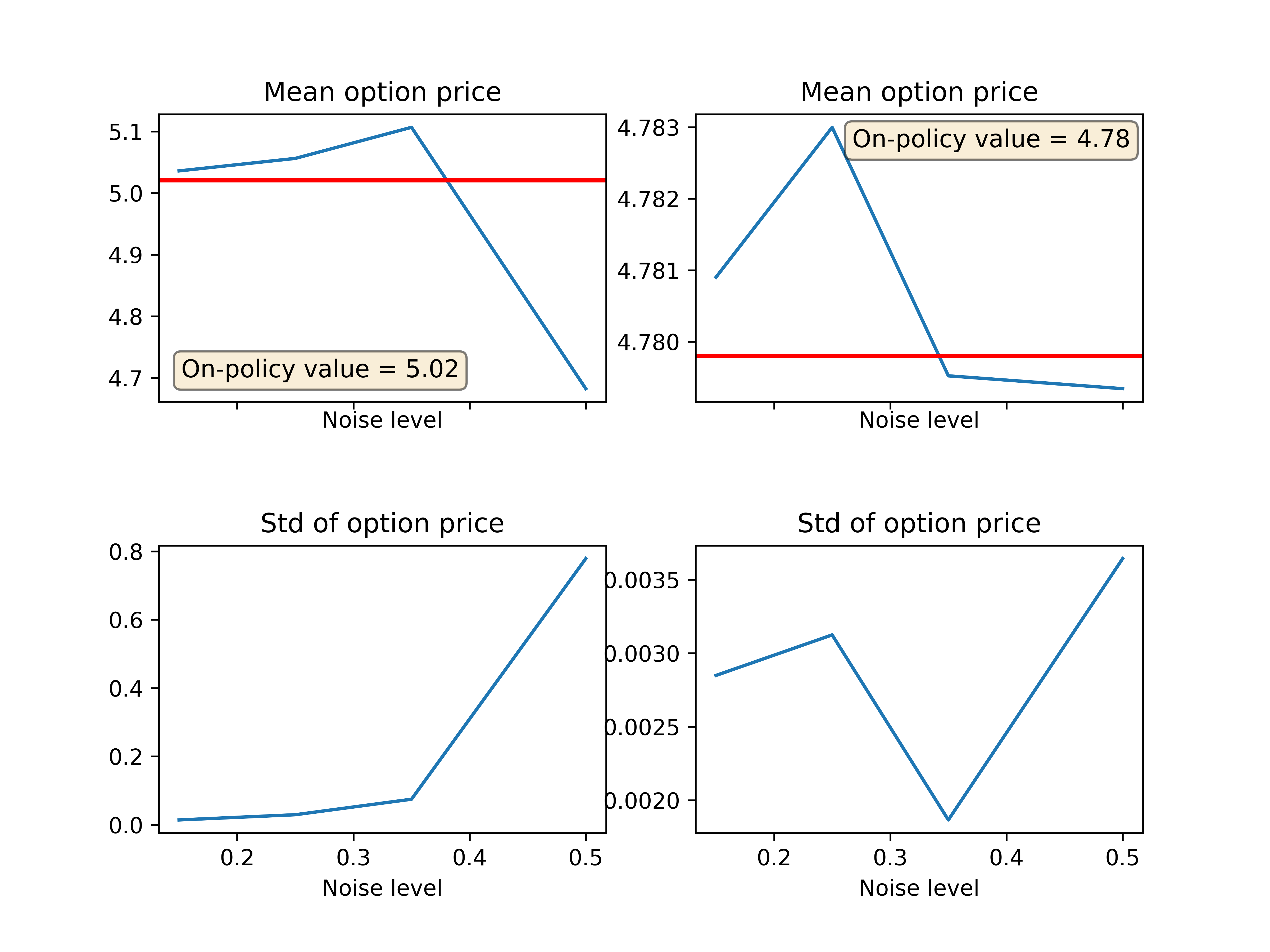}
\caption{Means and standard deviations of option prices obtained with {\it off-policy} FQI learning with data obtained by randomization of DP optimal actions by multiplying each optimal action by a uniform random variable in the interval $ [1- \eta, 1 + \eta] $ for $ \eta = [0.15, 0.25, 0.35, 0.5] $, with 5 scenarios for each value, and 2 MC runs. Horizontal red lines show values obtained with {\it on-policy} learning corresponding to $ \eta = 0 $.} 
\label{fig:Option_price_vs_noise_level}
\end{center}
\end{figure}

\begin{figure}[ht]
\begin{center}
\includegraphics[
width=140mm,
height=120mm]{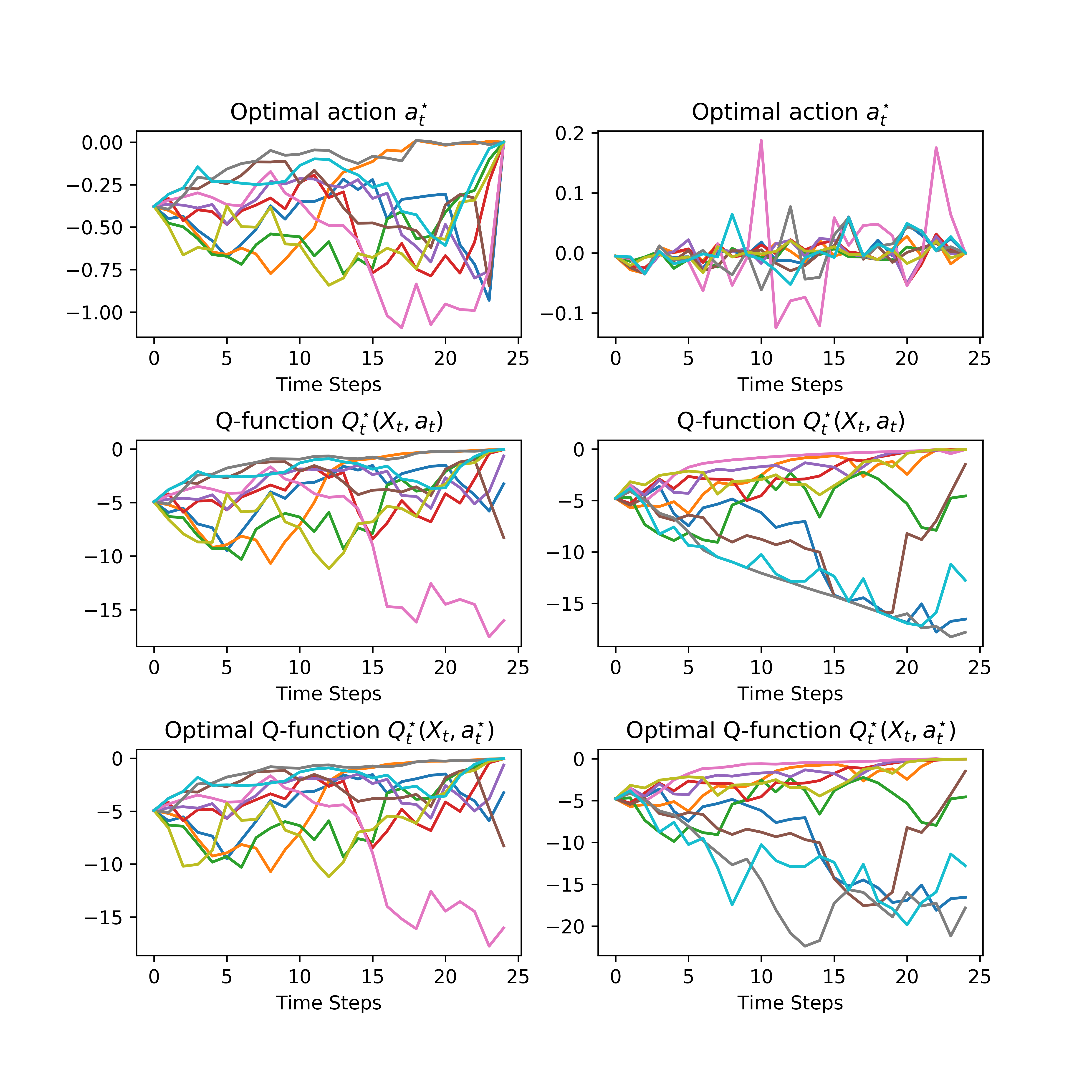}
\caption{RL solution (Fitted Q Iteration) for {\it off-policy} learning with noise parameter $ \eta = 0.5 $ for the ATM put option on a sub-set of MC paths for two MC runs.  } 
\label{fig:QLBS_FQI_off_policy_summary_ATM_eta_50}
\end{center}
\end{figure}

In the next set of experiments we deal with {\it off-policy} learning. To make off-policy data, we multiply, at each time step, optimal hedges computed by the DP solution of the model by a random uniform number in the interval $ [1- \eta, 1 + \eta] $ where $ 0 < \eta < 1 $ is a parameter controlling the noise level in the data. We will consider the values of $ \eta = [0.15, 0.25, 0.35, 0.5] $ to test the noise tolerance of our algorithms. Rewards corresponding to these sub-optimal actions are obtained using Eq.(\ref{one-step-reward}). In Fig.~\ref{fig:Option_price_vs_noise_level} we show results obtained for off-policy learning with 5 different scenarios of sub-optimal actions. Note that while some non-monotonicity in these graphs is due to a low number of scenarios, we note that the impact of sub-optimality of actions in recorded data is rather mild, at least for a moderate level of noise in actions. This is as expected as long as Fitted Q Iteration is an {\it off-policy} algorithm. This implies that when dataset is large enough, the QLBS model can learn even from data with purely random actions, and, in particular, learn the BSM model itself, if the world is lognormal \cite{IH_2017}.
Results of two MC runs for off-policy learning with the noise parameter $ \eta = 0.5 $ with Fitted Q Iteration algorithm  are shown in  
Fig.~\ref{fig:QLBS_FQI_off_policy_summary_ATM_eta_50}.

\section{Option portfolios}
\label{sect:Option_portfolios}

While above and in \cite{IH_2017} we looked at the problem of hedging and pricing of a single European option by an option seller that {\it does not} have any pre-existing option portfolio, here we outline a simple generalization to the case when the option seller {\it does} have such pre-existing option portfolio, or alternatively if she wants to sell a few options simultaneously\footnote{The context of this section was previously presented in a separate 
note "Relative Option Pricing in the QLBS Model" (https://papers.ssrn.com/sol3/papers.cfm?abstract\_id=3090608).}. 

In this case,  she needs to worry about 
{\it consistency} of pricing and hedging of {\it all} options in her new portfolio. In other words, she has to solve the dreaded {\it volatility smile problem} for her particular portfolio. Here we will show how she can do it in a {\it worry-free} way using the QLBS model.

Assume the option seller has a pre-existing portfolio of $ K $ options with market prices $ C_1, \ldots, C_K $. 
All these options reference an underlying state vector (market) $ {\bf X}_t $ which can be high-dimensional such that each particular option $ C_i $ with $ i = 1, \ldots, K $ references only one or a few components of market state $ {\bf X}_t $. 

Alternatively, we can add vanilla option prices as components of the market state $ {\bf X}_t  $.  In this case, our dynamic replicating portfolio would include vanilla options, along with underlying stocks. Such hedging portfolio would provide a dynamic generalization of static option hedging for exotics introduced by Carr {\it et. al.} \cite{Carr}.  

We assume that we have a historical dataset $ \mathcal{ F }$ that includes $ N $ observations of trajectories of tuples of vector-valued market factors, actions (hedges), and rewards (compare with Eq.(\ref{F_t_RL})):
\beq
\label{F_n}
\mathcal{F}_t^{(n)} = \left\{  \left( {\bf X}_t^{(n)}, {\bf a}_t^{(n)}, {\bf R}_t^{(n)}, {\bf X}_{t+1}^{(n)} 
\right) \right\}_{t=0}^{T-1} \, ,  \; \; n = 1, \ldots, N 
\eeq  
Now assume the option seller wants to add to this pre-existing portfolio another (exotic) option $ C_e $ (or alternatively, she wants to sell a portfolio of options $ C_1, \ldots, C_K, C_e $). Depending on whether the exotic option $ C_e $ was traded before in the market or not, there are two possible scenarios. We will look at them one by one.

In the first case, the exotic option $ C_e $ was previously traded in the market (by the seller herself, or by someone else). 
As long as its deltas and related P\&L impacts marked by a trading desk are available, we can simply extend vectors of 
actions $   {\bf a}_t^{(n)} $ and rewards $  {\bf R}_t^{(n)} $ in Eq.(\ref{F_n}), and then proceed with the FQI algorithm of Sect.~\ref{sect:FQI} (or with the IRL algorithm of Sect.~\ref{sect:IRL}, if rewards are not available). The outputs of the algorithm will be optimal price $ P_t $ of the whole option portfolio, plus optimal hedges for all options in the portfolio. Note that as long as FQI is an {\it off-policy} algorithm, it is quite forgiving to human or model errors: deltas in the data should not even be perfectly mutually consistent (see single-option examples in the previous section). But of course, the more consistency in the data, the less data is needed to learn an optimal portfolio price $ P_t $.

Once the optimal time-zero value  $ P_0 $ of the total portfolio $ C_1, \ldots, C_K, C_e $ is computed, a market-consistent price for the exotic option is simply given by a subtraction:
\beq
\label{subtraction}
C_e = P_0 - \sum_{i=1}^{K} C_i 
\eeq  
Note that by construction, the price $ C_e $ is consistent with {\it all} option prices $ C_1, \ldots, C_K $ and {\it all} their hedges, 
to the extent they are consistent between themselves (again, this is because Q-Learning is an off-policy algorithm).

Now consider a different case, when the exotic option  $ C_e $  was {\it not} previously traded in the market, and therefore there are 
no available historical hedges for this option. This can be handled by the QLBS model in essentially the same way as in the previous case. Again, because Q-Learning is an {\it off-policy} algorithm, it means that a delta and a reward of a {\it proxy} option 
$ C_e^{'} $ (that {\it was} traded before) to 
$ C_e $ could be used in the scheme just described in lieu of their actual values for option $ C_e $. Consistently with a common sense, this will just slow down the learning, so that more data would be needed to compute the optimal price and hedge for the exotic $ C_e $.  
On the other hand, the closer the traded proxy $ C_e^{'} $ to the actual exotic $ C_e $ the option seller wants to hedge and price, the 
more it helps the algorithm on the data demand side. Finally, when rewards for the $ C_e^{'} $ are not available, we can use the IRL methods 
of Sect.~\ref{sect:IRL}.

\section{Summary}
\label{sect:Summary}

In this paper, we have provided further extensions of the QLBS model developed in \cite{IH_2017}
for RL-based, data-driven and model-independent option pricing, including some topics for  "NuQLear" (Numerical Q-Learning) experimentations with the model. In particular, we have checked the convergence of the DP and RL solutions of the model to the BSM results in the limit $ \lambda \rightarrow 0 $. 

We looked into both {\it on-policy} and {\it off-policy} RL for option pricing, and showed that Fitted Q Iteration (FQI)  provides a reasonable level of noise tolerance with respect to possible sub-optimality of observed actions in our model, which is in agreement with general properties of Q-Learning being an {\it off-policy} algorithm. This makes the QLBS model capable of learning to hedge and price even when traders' actions (re-hedges) are sub-optimal or not mutually consistent for different time steps, or, in a portfolio context, between different options.  

We formulated an Inverse Reinforcement Learning (IRL) approach for the QLBS model, and showed that when the Markowitz risk aversion parameter
$ \lambda $ is {\it known}, the IRL and RL algorithms produce identical results, by construction. On the other hand, when $ \lambda $ is {\it unknown}, 
it can be separately estimated using Maximum Entropy (MaxEnt) IRL \cite{Ziebart_2008} applied to one-step transitions as in  
\cite{IRL_marketing}.  While this does {\it not} guarantee identical results between the RL and IRL solutions of the QLBS model, this can be assured again by using G-Learning \cite{G-Learning} instead of Q-Learning in the RL solution of the model.

Finally, we outlined how the QLBS model can be used in the context of option portfolios. By relying on Q-Learning and Fitted Q Iteration, which are
{\it model-free} methods, the QLBS model provides its own, data-driven and model independent solution to the (in)famous volatility smile problem of the Black-Scholes model.
While fitting the volatility smile and pricing options consistently with the smile is the {\it main objective} of Mathematical Finance option pricing models, 
this is just a {\it by-product} for the QLBS model. This is because the latter is distribution-free, and is therefore capable of adjusting to {\it any} smile (a set of market quotes for vanilla options).  As was emphasized in the Introduction and in \cite{IH_2017}, the {\it main} 
difference between  all continuous-time option pricing models of Mathematical Finance (including the BSM model and its various local and stochastic volatility extensions, jump-diffusion models, etc.) and the QLBS model is that while the former try to "match the market", they remain clueless about the expected risk in option positions, while the QLBS model makes the {\it risk-return} analysis of option replicating portfolios the {\it main focus} of option hedging and pricing, similarly to how such analysis is applied for stocks in the classical Markowitz portfolio theory \cite{Markowitz}. 


%
%

\end{document}